\documentclass[a4paper,11pt]{article}
\usepackage{jheppub} % for details on the use of the package, please see the JHEP-author-manual
\usepackage[T1]{fontenc} % if needed
\usepackage{multirow}
\usepackage{color}
\usepackage{ulem}
\usepackage{rotating}
\usepackage{booktabs}
\usepackage{dcolumn}
\usepackage{array} 
\usepackage{multirow}
\usepackage{lineno}
\usepackage{siunitx}
\usepackage{amsmath}
\usepackage{bm}
\usepackage{mathrsfs}
\usepackage{indentfirst}

\DeclareSIUnit{\bmm}{\bm{m}}

\DeclareSIUnit{\clight}{\textnormal{\textit{c}}}

\newcolumntype{d}{D{.}{.}{-1}}
\newcolumntype{e}{D{.}{.}{8}}
\newcolumntype{f}{D{.}{.}{18}}
\newcolumntype{h}{D{.}{.}{13}}
\newcolumntype{g}{D{.}{.}{12}}
\newcommand{\EE}{e^+e^-}
\newcommand{\ar}{\rightarrow}
\newcommand{\SSB}{\Sigma^+\bar{\Sigma}^-}
\newcommand{\ssb}{\Sigma^{0}\bar\Sigma^{0}}
\newcommand{\chicJ}{\chi_{cJ}}

\title{\protect\boldmath Measurement of $\Sigma^+$ transverse polarization in $e^+e^-$ collisions at $\sqrt{s} = 3.68-3.71$ GeV}

\collaborationImg{\includegraphics[width=0.25\textwidth]{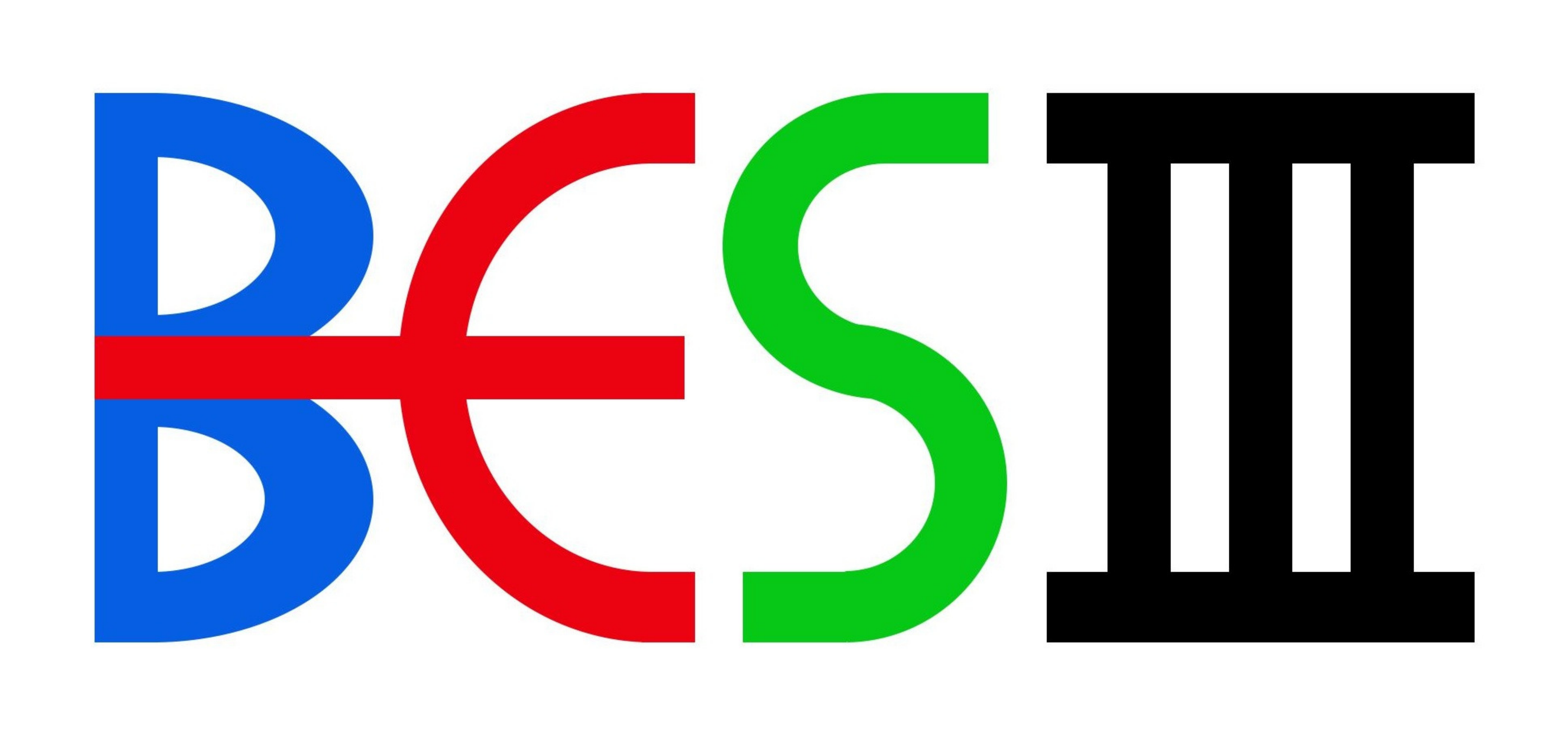}}
\collaboration{The BESIII Collaboration}

\emailAdd{besiii-publications@ihep.ac.cn}

\begin{document} 

\abstract{Using $e^+e^-$ collision data collected with the BESIII
  detector at seven energy points ranging from 3.68 to 3.71 GeV and
  corresponding to an integrated luminosity of $652.1~{\rm pb^{-1}}$,
  we present an energy-dependent measurement of the transverse
  polarization, relative phase and modulus ratio of the electromagnetic form factors
  of the $\Sigma^+$ hyperon in the $e^+e^- \to \Sigma^+
  \bar{\Sigma}^-$ reaction. At each energy point, no evident polarization is observed.
  These results are helpful to understand
  the production mechanism of the $\Sigma^+$-$\bar\Sigma^-$ pairs. }

% \begin{document}
%\include{def-com}
\maketitle
\flushbottom
%\linenumbers

\section{Introduction} \label{sec:intro}
\noindent
As early as 1960, the significance of baryon structure was widely
acknowledged~\cite{Cabibbo:1961sz}. However, at present, understanding
the structure of baryons remains a great challenge. Apart from protons
and neutrons, knowledge has been relatively scarce. The proton, being
a stable particle, allows for the determination of the space-like
electromagnetic form factors (EMFFs) via elastic electron-proton
scattering~\cite{Perdrisat:2006hj}. More recently, $e^+e^-$collisions,
resulting in the production of baryon pairs, provide an ideal
experimental framework to study baryon structure. This is because the
$e^+e^-$ process offers access to the time-like EMFFs via virtual photon
production, thereby facilitating the quantitative assessment of
baryonic electromagnetic structure. Experimentally accessible
time-like EMFFs are intimately connected with more intuitive
space-like quantities, such as charge and magnetization densities,
through the dispersion relation~\cite{Belushkin:2006qa,
  Yan:2023yff}. Baryon pairs with a spin of 1/2 produced by the $e^+e^-$
process via a virtual photon are elegantly described by two distinct
EMFFs: the electric form factor $G_E$ and magnetic form factor
$G_M$~\cite{Huang:2021xte, Aubert:2007uf}. These form factors are
functions of the square of the four-momentum transfer, denoted as
$q^2$. In the time-like region, where $q^2 > 0$, the EMFFs exhibit
non-zero imaginary components. Consequently, when $G_E$ and $G_M$ are
different, they give rise to a relative phase denoted as $\Delta\Phi =
\Phi_E - \Phi_M$, where $\Phi_E$ and $\Phi_M$ are the phases of $G_E$
and $G_M$, respectively.

In accordance with the Phragmén-Lindelöf
theorem~\cite{titchmarsh1939theory}, as $q^2\to\infty$, the asymptotic
behavior of the time-like EMFFs can be obtained from their
corresponding space-like counterparts. In the space-like region, the
EMFFs are assumed to be real. Consequently, as $q^2\to\infty$, the
time-like EMFFs must also be real, and the relative phase of the EMFFs
should approach an integer multiple of $\pi$. In the case where $q^2$
does not approach the limit, the phase can be any value, which causes the 
final state to be polarized, even if the
initial state is unpolarized~\cite{Dubnickova:1992ii}. This allows the
determination of the EMFFs, as the polarization and cross-section are
functions of form factors themselves~\cite{Faldt:2017kgy,Cao:2021asd, Dai:2023vsw, 
Yan:2023yff, Chen:2024luh}.  Recent
results from BESIII~\cite{BESIII:2020uqk},
BaBar~\cite{Aubert:2007uf} and Belle~\cite{Belle:2022dvb},
focusing solely on measuring the effective form factor, have
reported consistent results. Additionally, the CLEO collaboration has
made significant progress in determining the cross sections and the
EMFFs of various baryon pairs ($p$, $\Lambda$, $\Sigma^0$, $\Sigma^+$,
$\Xi^0$, $\Xi^-$, and $\Omega^-$) \cite{Dobbs:2014ifa,
  Dobbs:2017}. Their conclusions regarding EMFFs and di-quark
correlations~\cite{Jafee:2005,Jafee:2003} rely on the assumption that
one-photon exchange dominates the production process and that
charmonia contributions are negligible.  Furthermore, the BESIII
collaboration has measured the cross sections for several baryon pairs
($\Lambda$, $\Sigma^0$, $\Xi^-$, $\Sigma^\pm$, and $\Xi^0$) near the
production threshold~\cite{BESIII:2017hyw, BESIII:2020uqk, BESIII:2020ktn, BESIII:2021aer, BESIII:2021rkn, BESIII:2022kzc} and above open
charm threshold~\cite{BESIII:2019cuv, BESIII:2021ccp, BESIII:2023rse, BESIII:2024umc, BESIII:2024ogz}. However,
experimental measurements of the relative phase between $G_E$ and
$G_M$ are still relatively scarce.

For spin 1/2 baryon pairs produced by the $e^+e^-$ process via vector
charmonia, where the above statements concerning EMFFs also apply, the
formalism is described in ref.~\cite{Faldt:2017kgy}. In these cases,
the amplitudes include the EM-\textit{psionic} form factors,
$G^{\Psi}_E$ and $G^{\Psi}_M$~\cite{BESIII:2021cvv,BESIII:2023euh}. While the
EM-\textit{psionic} form factors describe a different process, the
form of the hadron current matrix element
for the charmonia process is the same as that
for the virtual photon
one~\cite{Faldt:2017kgy,Faldt:EPJ}.

Previous studies often neglected the polarization effects of hyperons
\cite{ Ablikim:2016iym, Ablikim:2016iym-01,
  Ablikim:2016iym-02, Ablikim:2016iym-03, Wang:2018kdh, Wang:2021lfq}. 
 However, $\Sigma^+$
polarization was recently observed and measured in the $e^+e^-\to
J/\psi$, $\psi(3686)\to\Sigma^+\bar{\Sigma}^-$ processes by the BESIII
collaboration~\cite{BESIII:2020fqg,BESIII:2023sgt}. The results not only reveal a
non-zero relative phase but also demonstrate that the phase
changes sign at the mass of the $\psi(3686)$ resonance compared to the
value measured at the $J/\psi$ resonance. Subsequently, the
polarization effects of $\Lambda$ and $\Xi^-$ were also observed and
measured by the BESIII collaboration. The $\Lambda$ polarization was
observed in the $e^+e^-\to\Lambda\bar{\Lambda}$ process at the
$J/\psi$, $\psi(3770)$, and off-resonance
regions~\cite{
BESIII:2022yprl, Ablikim:2018zay,Wang:2023trb}. Additionally, a non-zero polarization has been
observed for $\Xi^-$ in the $\EE\ar J/\psi$,
$\psi(3686)\ar\Xi^-\bar{\Xi}^+$ processes~\cite{BESIII:2021ypr,BESIII:2022_xi_psip,Liu:2023xhg}. However, for $\Xi^0$, completely different
effects were observed in processes $\EE\ar J/\psi \ar
\Xi^0\bar{\Xi}^0$~\cite{BESIII:2023drj} and $\EE\ar \psi(3686) \ar
\Xi^0\bar{\Xi}^0$~\cite{BESIII:2023lkg}.

The data samples used in this analysis correspond to an integrated
luminosity of 652.1 pb$^{-1}$, collected at the center-of-mass (CM)
energies of
 $\sqrt{s} =$ 3.682, 3.683, 3.684, 3.685, 3.687, 3.691, and 3.710 GeV
 with the BESIII detector~\cite{Ablikim:2009aa} in symmetric
 $e^+e^-$ collisions provided by the BEPCII storage
 ring~\cite{BEPCII}.
The above energy points around $\psi(3686)$ resonance, which are used to study the
 energy dependence of
 $\Delta\Phi$,
 are particularly intriguing, as
 the production occurs through a combination of one-photon
 exchange~\cite{BESIII:2019nep}, mixed with the $\psi(3686)$
 resonance~\cite{BESIII:2021cvv}, and resonance
 dominance~\cite{Ablikim:2018zay, BESIII:2022yprl}. Also, it can enrich the experimental information of form factors measurement 
 in higher energy region~\cite{Ramalho:2024wxp}. Since the energy
 range studied here includes both the virtual photon and
 vector charmonium processes, $G_E^{\gamma/\Psi}$ and $G_M^{\gamma/\Psi}$ are used to represent the form factors in the
 following, and the methods used in the following apply to both
 processes.

\section{BESIII detector and Monte Carlo simulation}
\noindent
The BESIII detector~\cite{Ablikim:2009aa} records symmetric $e^+e^-$
collisions provided by the BEPCII storage ring~\cite{BEPCII} in the CM
energy ranging from 1.84 to \SI{4.95}{GeV}, with a peak luminosity of
\SI{1e33}{\per\centi\meter\squared\per\second} achieved at $\sqrt{s}
=$ 3.773 {GeV}. BESIII has collected large data samples in this energy
region~\cite{Ablikim:2019hff, EcmsMea, EventFilter}. The cylindrical
core of the BESIII detector covers 93\% of the full solid angle and
consists of a helium-based multilayer drift chamber~(MDC), a plastic
scintillator barrel and multigap resistive plate chamber end cap
time-of-flight system~(TOF), and a CsI(Tl) electromagnetic
calorimeter~(EMC), all enclosed in a superconducting solenoidal magnet
providing a \SI{1.0}{T} magnetic field. The solenoid is supported by
an octagonal flux-return yoke with resistive plate counter muon
identification modules interleaved with steel. The charged-particle
momentum resolution at \SI{1}{GeV/\clight} is $0.5\%$, and the ${\rm
d}E/{\rm d}x$ resolution is $6\%$ for electrons from Bhabha
scattering. The EMC measures photon energies with a resolution of
$2.5\%$ ($5\%$) at \SI{1}{GeV} in the barrel (end cap) region. The
time resolution in the TOF barrel region is \SI{68}{ps}, while that in
the end cap region is \SI{60}{ps}~\cite{etof1,etof2,etof3}.

To evaluate detection efficiencies and estimate backgrounds, simulated
data samples are produced using {\sc geant4}-based Monte Carlo (MC)
software~\cite{GEANT4}, which incorporates the geometric description
of the BESIII detector~\cite{Huang:2022wuo} as well as the detector
response. The simulation models the beam energy spread and
initial-state radiation (ISR) effect in the $e^+e^-$ annihilation
process with {\sc kkmc}~\cite{KKMC}. The detection efficiency for the
$\EE\to\SSB$ process is determined through MC simulations. For each of
the 7 energy points ranging from 3.68 to 3.71 GeV, a sample of
100,000 events is simulated with a uniform phase space (PHSP)
distribution. The $\Sigma^+(\bar{\Sigma}^-)$ baryon and its subsequent
decays are simulated using {\sc evtgen}~\cite{evtgen2,EVTGEN} with a
PHSP model.

\section{Event selection}
\noindent
A double-tag technique is employed in the event selection
involving the decays of $\Sigma^+\to p\pi^0$ and
$\bar{\Sigma}^-\to\bar{p}\pi^0$ with the subsequent decay
$\pi^0\ar\gamma\gamma$. Hence, in the final state, there are two
charged particles, a proton and an anti-proton, along
with four photons utilized for the reconstruction of two
$\pi^0$s. 
Consequently, suitable candidates must meet the following
event selection criteria.

%Tracking and PID

Charged tracks detected in the MDC are required to be within a polar
angle ($\theta$) range of $\vert\!\cos\theta\vert < 0.93$, where
$\theta$ is defined with respect to the $z$-axis, which is the
symmetry axis of the MDC. At least two oppositely charged tracks,
which must be well reconstructed in the MDC with good helix fits, are
required. For the distance $\rm V_{xy}$ between the vertex 
and the xy plane, and the distance $\rm V_z$ between the vertex and 
the z-axis, both the signal and background distributions are around 0. To avoid efficiency loss, there is no vertex requirement. 
Since (anti-)protons from
$\Sigma^+(\bar{\Sigma}^-)$ decays can be distinguished from other
charged particles by requiring the momentum to be greater than
$0.5~{\rm GeV}/c$, there is no additional particle identification
(PID) requirement.

%gamma selection

Showers deposited in the EMC are used to reconstruct $\pi^0$s. The
deposited energy of each shower must be more than 25~MeV in the barrel
region ($\vert\!\cos\theta\vert< 0.80$) and more than 50~MeV in the
end cap region ($0.86 <\vert\!\cos\theta\vert< 0.92$). To suppress
electronic noise and showers unrelated to the event, the difference
between the EMC time and the event start time is required to be within
[0, 700] ns. After selection, at least four photons are required.

%Sigma reconstruction

\begin{figure}[!hbpt] \begin{center}
\includegraphics[width=0.8\textwidth]{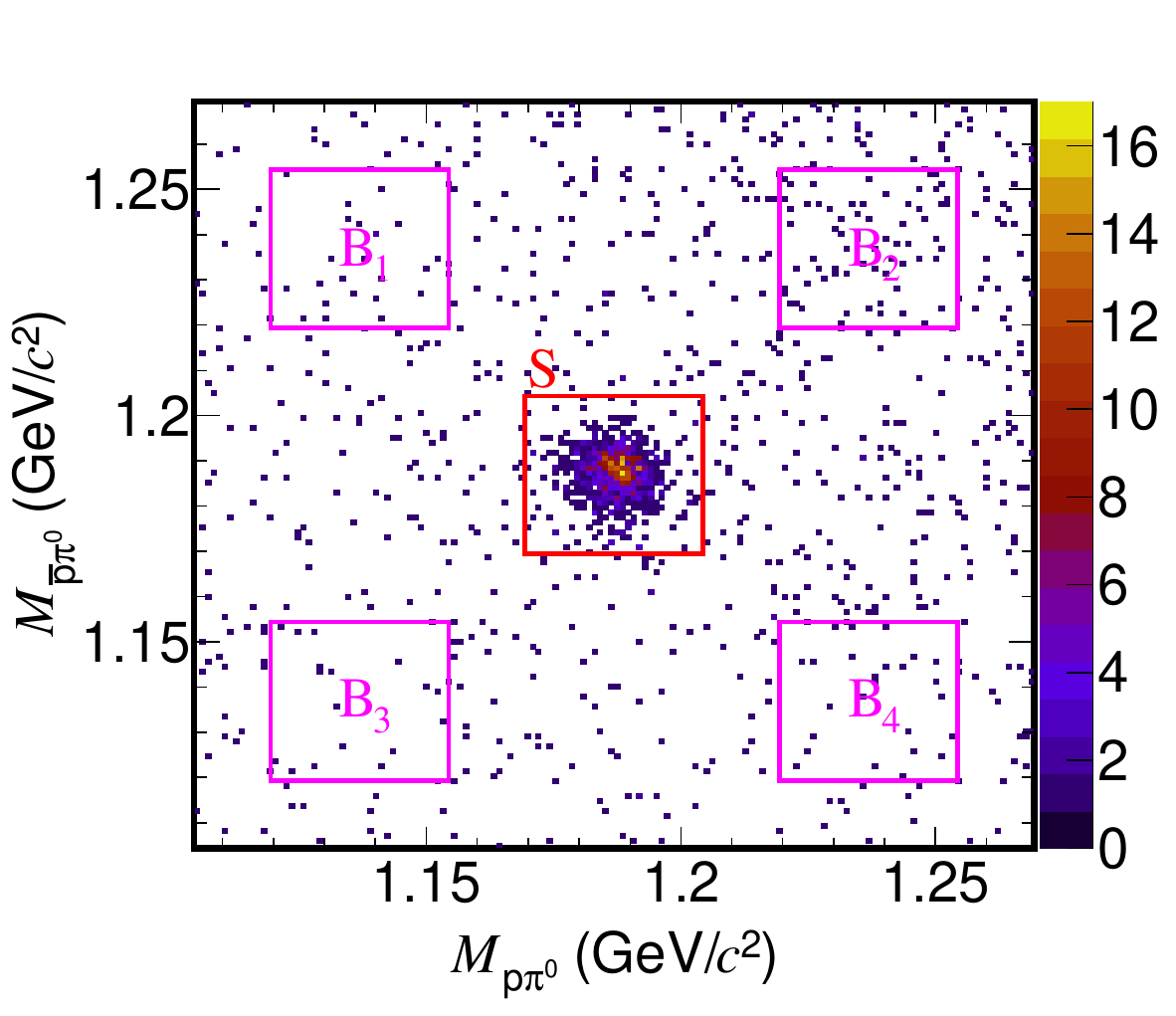} \end{center}
\caption{ Two-dimensional distribution of $M_{\bar{p}\pi^0}$ versus
$M_{p\pi^0}$ of the accepted candidates summed over all energy points,
where the red square marked with $S$ indicates the signal region, and
the pink squares marked with $B_i~(i=1,2,3,4)$ show the selected
background regions.  } \label{Fig:SUM:DATA} \end{figure}

To select the correct combination of proton, anti-proton, and
$\pi^0\pi^0$ candidates, a six-constraint (6C) kinematic fit is
applied to all $p\bar{p}\gamma\gamma\gamma\gamma$ combinations in each
event. The 6C kinematic fit conserves energy and momentum while
constraining the invariant mass of photon combinations to the known
$\pi^0$ mass~\cite{PDG2020}. The $p\bar{p}\pi^0\pi^0$ combination with
the smallest $\chi^2$ is retained, and $\chi_{\rm 6C}^2 < 100$ is
required to suppress the background. This requirement is determined by
analyzing the figure-of-merit (FOM) $N_{\rm sig}/\sqrt{N_{\rm
sig}+N_{\rm bkg}}$, where $N_{\rm sig}$ and $N_{\rm bkg}$ represent
the numbers of signal and background events, respectively, both based
on MC simulation.  To match $p$ and $\bar{p}$ with the correct
$\pi^0$, the $\SSB$ pair with the minimum difference
$\sqrt{(M_{p\pi^0}-m_{\Sigma^+})^2+(M_{\bar{p}\pi^0}-m_{\bar{\Sigma}^-})^2}$
is selected for further analysis. Here, $M_{p\pi^0(\bar{p}\pi^0)}$ is
the invariant mass of the $p\pi^0(\bar{p}\pi^0)$ combination, and
$m_{\Sigma^+ (\bar{\Sigma}^-)}$ is the known $\Sigma^+$
mass~\cite{PDG2020}. To eliminate the primary background originating
from the $\EE\ar\pi^0\pi^0J/\psi\ar p\bar{p}\pi^0\pi^0$ process, the
veto requirement $|M_{\pi^0\pi^0}^{\rm Recoil} - m_{J/\psi}| > 15$
MeV/$c^2$ is applied. Here, $M_{\pi^0\pi^0}^{\rm Recoil}$ is the
recoil mass of the $\pi^0\pi^0$ combination and $m_{J/\psi}$ is the
mass of $J/\psi$~\cite{PDG2020}. This requirement is also determined
by FOM optimization.  Figure~\ref{Fig:SUM:DATA} shows the distribution
of $M_{\bar{p}\pi^0}$ versus $M_{p\pi^0}$ of the accepted candidates
summed over all energy points. The $M_{p\pi^0(\bar{p}\pi^0)}$
candidates are required to be within the range of $[m_{\Sigma^+} -
4\sigma, m_{\Sigma^+} + 3\sigma]$ MeV/$c^2$, denoted by $S$ in
figure~\ref{Fig:SUM:DATA}, where the resolution $\sigma$ is determined
by a one-dimensional fit with the Crystal-Ball
function~\cite{Oreglia:1980cs}. Due to the energy leakage of the
photon in the EMC, the signal shape is asymmetric.

After applying the event selection criteria to the data, the remaining
background mainly comes from non-$\Sigma^+(\bar{\Sigma}^-)$ events,
such as the non-resonant process $\EE\ar\pi^0\pi^0p\bar{p}$. The number of
background events is estimated using the sideband method, {\it i.e.} 
$\sum^{4}_{i=1}B_{i}/4$ for the $M_{p\pi^0}$ and $M_{\bar{p}\pi^0}$
windows, where $i$ runs over the four regions shown in
figure~\ref{Fig:SUM:DATA}, and the exact ranges are defined as
$B_{1}$: $M_{p\pi^0}~\in$ [1.119, 1.154] GeV/$c^2$ and $M_{\bar{p}\pi^0}~\in$ [1.219, 1.254] GeV/$c^2$,
$B_{2}$: $M_{p\pi^0}~\in$ [1.219, 1.254] GeV/$c^2$ and $M_{\bar{p}\pi^0}~\in$ [1.219, 1.254] GeV/$c^2$,
$B_{3}$: $M_{p\pi^0}~\in$ [1.119, 1.154] GeV/$c^2$ and $M_{\bar{p}\pi^0}~\in$ [1.119, 1.154] GeV/$c^2$,
$B_{4}$: $M_{p\pi^0}~\in$ [1.219, 1.254] GeV/$c^2$ and $M_{\bar{p}\pi^0}~\in$ [1.119, 1.154] GeV/$c^2$.

The signal yield combining all energy points is $898\pm30$
(stat.) events. The numbers of signal events for each energy point
are listed in table~\ref{summary}. The number of background events
estimated with the sideband method is $36\pm3$
(stat.), which is a background level of 3.85\%.

\section{\boldmath $\Sigma^+$ polarization}
\begin{figure}[!htbp] \centering
\includegraphics[width=0.95\textwidth]{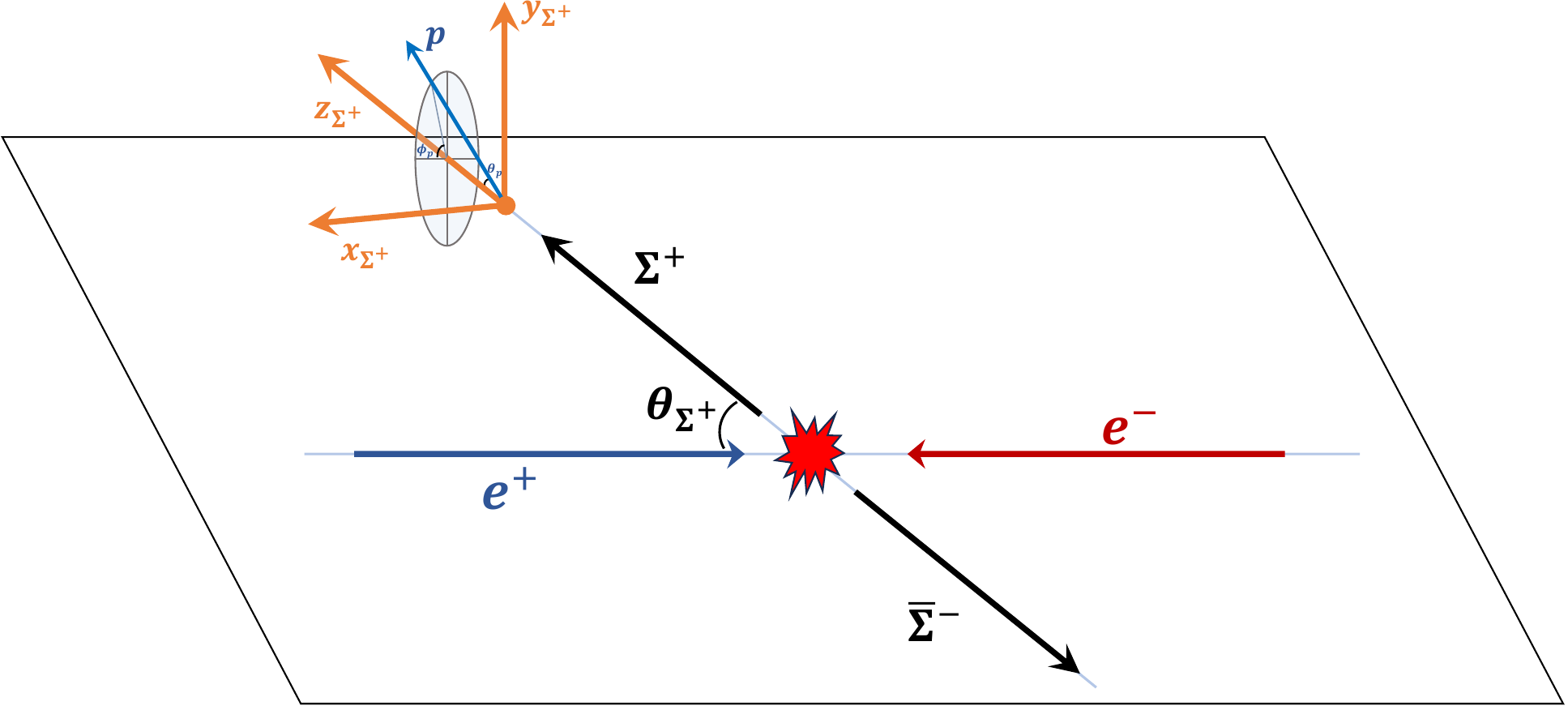} \caption{
Definition of the coordinate system describing the
  $\EE\ar\gamma^*/\Psi\ar\SSB\ar p\bar{p}\pi^0\pi^0$ reaction. The
  $\Sigma^+$ particle is emitted along the $z_{\Sigma^+}$ axis
  direction, and the $\bar{\Sigma}^-$ in the opposite direction. The
  $y_{\Sigma^+}$ axis is perpendicular to the plane of $\Sigma^+$ and
  $e^{-}$, and the $x_{\Sigma^+}$ axis is defined by a right-handed
  coordinate system. The $\Sigma^+$ decay product, the proton, is
  measured in this coordinate system.}  \label{fig:helicity_frame}
  \end{figure}

\noindent
The exclusive process
$\EE\ar\gamma^*/\Psi\ar\SSB\ar p\bar{p}\pi^0\pi^0$ can be fully
described by the $\Sigma^+$ scattering angle, $\theta_{\Sigma^+}$, in
the CM system of the $e^+e^-$ reaction and the $p$ and ~$\bar{p}$
directions, $\boldsymbol{\hat{n}_{1}}$ and $\boldsymbol{\hat{n}_{2}}$,
respectively, in the rest frames of their parent particles. Here
$\gamma^*/\Psi$ indicates that the process $\EE\ar\SSB$ occurs via a
pure EM process or a $\psi$ resonance. The components of these vectors
are expressed using a coordinate system ($x_{\Sigma^+},y_{\Sigma^+},
z_{\Sigma^+}$) as shown in
figure~\ref{fig:helicity_frame}. A right-handed system for each
hyperon decay is defined with the $z$ axis along the $\Sigma^+$
momentum $\textbf{p}_{\Sigma^+} = - \textbf{p}_{\bar{\Sigma}^-} =
\textbf{p}$ in the CM system. The $y$ axis is taken as the normal to
the scattering plane, $\textbf{p}_{\Sigma^+} \times \textbf{k}_{e^{-}}$, 
where $\textbf{k}_{e^{-}} = -
\textbf{k}_{e^{+}} = \textbf{k}$ is the electron beam momentum in the
CM system.  

The joint decay angular distribution of the process
$\EE\ar\gamma^*/\Psi\ar\SSB\ar p\bar{p}\pi^{0}\pi^{0}$, involving
spin-entangled $\SSB$, is expressed
as~\cite{Faldt:2017kgy}

\begin{align}
\label{eq:tangles:abcd}
{\cal{W}}(\boldsymbol{\xi}; \boldsymbol{\Omega}) = & F_0(\xi) + \eta F_5(\xi) + \alpha_{\Sigma^+}\alpha_{\bar{\Sigma}^-} \nonumber\\
\times &[F_1(\xi) + \sqrt{1-\eta^2} \cos(\Delta\Phi)F_2(\xi) + \eta F_6(\xi)] \nonumber \\
+ &\sqrt{1-\eta^{2}}\sin(\Delta\Phi)[\alpha_{\Sigma^+}F_3(\xi) + \alpha_{\bar\Sigma^-}F_4(\xi)], 
\end{align}
where $\eta$ is the angular distribution parameter,
\boldsymbol{$\Omega$} $= (\eta, \Delta\Phi, \alpha_{\Sigma^+},
\alpha_{\bar{\Sigma}^-})$ represents the production and decay
parameters, the kinematic variables \boldsymbol{$\xi$} $=
(\theta_{\Sigma^+}, \boldsymbol{\hat{n}_1}, \boldsymbol{\hat{n}_2})$
describe the production and subsequent decay, and
$\alpha_{\Sigma^+(\bar{\Sigma}^-)}$ denotes the asymmetry parameter of the
$\Sigma^+(\bar{\Sigma}^-)\ar p\pi^0(\bar{p}\pi^0)$ decay. The angular
functions $F_i(\xi)~(i=0,1,...6)$ are described in detail in
ref.~\cite{Faldt:2017kgy}, and the definitions of five terms with angular distribution of final states are shown in eq.~(\ref{eq:tangles}).
\begin{align}
 \label{eq:tangles}
F_1 = &\sum_{i=1}^{N_k}(\sin^2\!\theta_{\Sigma^+} n^i_{1, x}n^i_{2, x} + \cos^2\!\theta_{\Sigma^+} n^i_{1, z}n^i_{2, z}),\nonumber\\
F_2 = &\sum_{i=1}^{N_k}\sin\!\theta_{\Sigma^+}\cos\!\theta_{\Sigma^+} (n^i_{1, x}n^i_{2, z} + n^i_{1, z}n^i_{2, x}),\nonumber\\
F_3 = &\sum_{i=1}^{N_k}\sin\!\theta_{\Sigma^+}\cos\!\theta_{\Sigma^+} n^i_{1, y},\\
F_4 = &\sum_{i=1}^{N_k}\sin\!\theta_{\Sigma^+}\cos\!\theta_{\Sigma^+} n^i_{2, y},\nonumber\\
F_6 = &\sum_{i=1}^{N_k}(n^i_{1, z}n^i_{2, z} - \sin^2\!\theta_{\Sigma^+} n^i_{1, y}n^i_{2, y}),\nonumber
\end{align}
where, $N_k$ is the number of events in the $k^{\rm th}$
$\cos\theta_{\Sigma^+}$ interval, $n_{1,j}~(n_{2,j})~(j=x,y,z)$
represents the component of vector $\bold{\hat{n}}_1$
($\bold{\hat{n}}_2$) in the coordinate system ($x_{\Sigma^+}$,
$y_{\Sigma^+}$, $z_{\Sigma^+}$), and $i$ is the index from 1 to
$N_k$.

The modulus ratio of the form factors, $R$~\cite{BESIII:2021cvv}, can be described by the angular
distribution parameter $\eta$. 
\begin{equation}%\label{FF05}
R= \frac{|G_E^{\gamma/\Psi}|}{|G_M^{\gamma/\Psi}|} = \sqrt{ \frac{\tau(1-\eta)}{1+\eta}},
%\label{eq:R}
\end{equation}
where $\tau = \frac{s}{4m_{\Sigma^+}^2}$ and $s$ is the square of the
CM energy. If the initial state is unpolarized, and the production
process is either strong or electromagnetic and hence
parity-conserving, then a non-zero polarization is only possible in
the transverse direction $y$. The polarization is given by~\cite{Faldt:2017kgy}
\begin{equation}
P_y=\frac{\sqrt{1-\eta^2}\sin\theta_{\Sigma^+}\cos\theta_{\Sigma^+}}{1+\eta\cos^2\theta_{\Sigma^+}}\sin(\Delta\Phi).
\label{eq:pol}
\end{equation}

To determine $\eta$ and $\Delta\Phi$, an unbinned maximum likelihood
fit is performed, where the decay parameters $\alpha_{\Sigma^+}$ and
$\alpha_{\bar{\Sigma}^-}$ are fixed to the values $-0.994$ and 0.994,
respectively, obtained from the average in ref.~\cite{BESIII:2020fqg},
assuming $CP$ conservation. In the fit, the likelihood function
$\mathscr{L}$ is constructed from the probability function,
${\cal{P}}({\boldsymbol{\xi}}_i)$, for event $i$ characterized by the
measured angles $\boldsymbol{\xi}_i$ \begin{equation} \mathscr{L} =
\prod_{i=1}^N {\cal{P}}({\boldsymbol{\xi}}_i, {\boldsymbol{\Omega}}) =
\prod_{i=1}^N {\cal{C}}{\cal{W}}({\boldsymbol{\xi}}_i,
{\boldsymbol{\Omega}})\epsilon(\boldsymbol{\xi}_i), \end{equation}
where $N$ is the number of events in the signal region, and
$\epsilon(\boldsymbol{\xi}_i)$ is the detection efficiency. For the
ISR effect at the higher energy points 3.691 and 3.710 GeV, MC studies
are performed where the input cross section for $\EE\ar\SSB$ for
calculating the ISR effect is taken from ref.~\cite{borncs}. The ISR
effect at these two energy points brings absolute differences of 0.02
and 0.04 rad for $\eta$ and $\Delta\Phi$, respectively, which are
negligible. The normalization factor
$\mathcal{C}=\frac{1}{N_\mathrm{MC}}\sum_{j=1}^{N_\mathrm{MC}}
{\cal{W}}({\boldsymbol{\xi}}^{j}, {\boldsymbol{\Omega}})$ is given by
the sum of the corresponding angular distribution function $\cal{W}$
using the accepted MC events, $N_\mathrm{MC}$, and the difference in
efficiency between data and MC simulations is taken into account as a
systematic uncertainty, described later. The minimization of the
function \begin{equation} \mathscr{S} = -\mathrm{ln}\mathscr{L}_{\rm
data}, \end{equation} is performed with RooFit~\cite{Roofit}. Here,
$\mathscr{L}_{\rm data}$ is the likelihood function of events selected
in the signal region.

\begin{figure}[!htbp]
  \centering
  \includegraphics[width=0.95\textwidth]{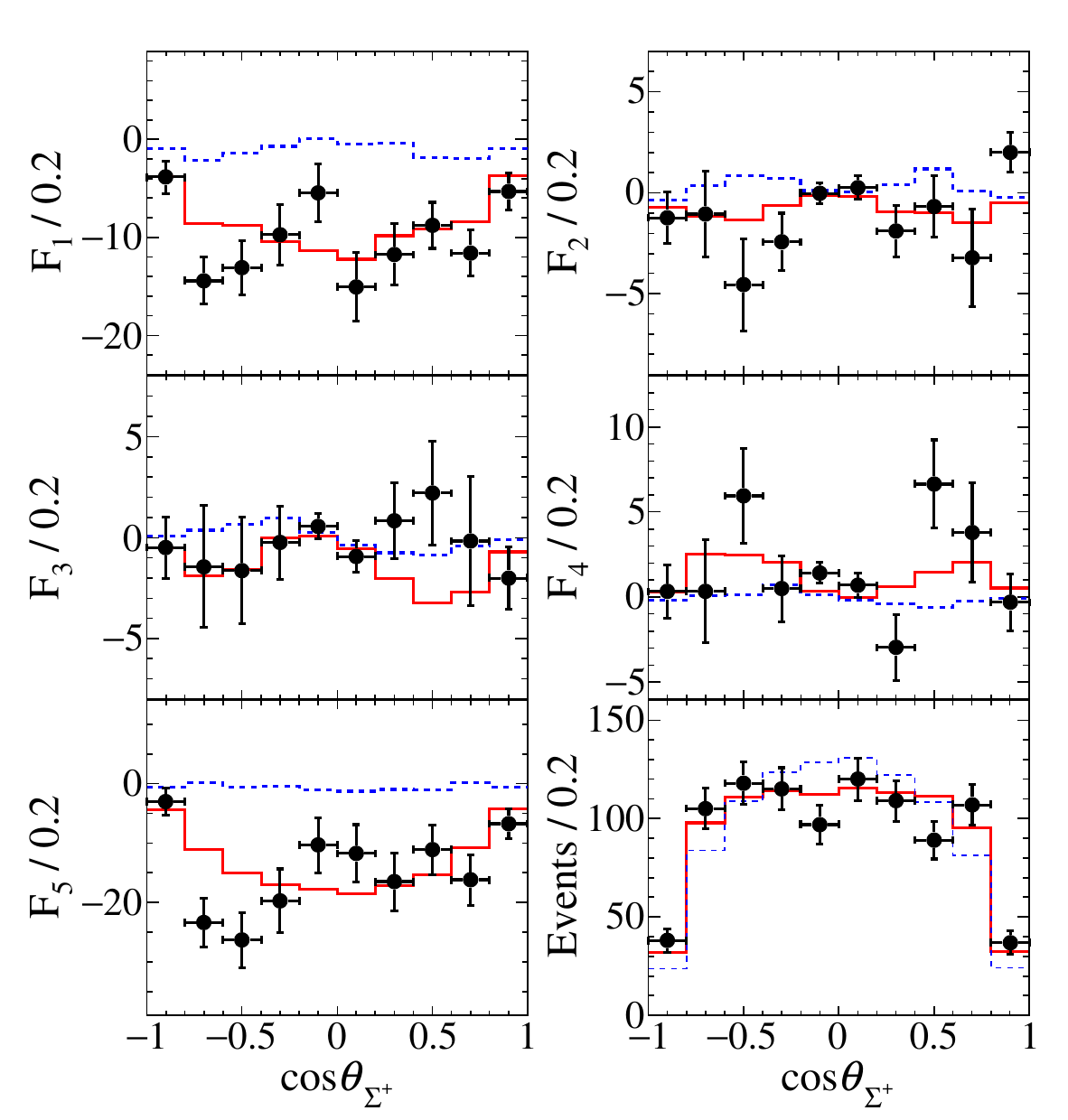}
  \caption{ Distributions of ${F}_{k} (k = 1, 2, 3, 4, 6)$ moments with
    respect to $\cos\theta_{\Sigma^+}$. The dots with error bars are data combined from all
    energy points, and the red solid lines are the weighted PHSP MC
    corrected by the results of global fit. The blue dashed lines
    represent the distributions of the simulated events evenly
    distributed in PHSP, without polarization.}
  \label{scatter_plot::llb:projections}
  \end{figure}

Figure~\ref{scatter_plot::llb:projections} shows the distributions of
the five ${F}_{k} (k = 1, 2, 3, 4, 6)$ moments \cite{Faldt:2017kgy}
with respect to $\cos\theta_{\Sigma^+}$, divided into 10
intervals, and the weighted PHSP MC results corrected by the global fit. 
The numerical fit results and the weighted average values are summarized in table~\ref{summary}. 
The weighted average can provide an overall evaluation of 
the $\Sigma^+$ polarization related to the production mechanism of the $\Sigma^+\bar\Sigma^-$ pair 
around this energy range. It is calculated by
$\frac{1}{w}\sum_{i=1}^{n}w_ix_i$~\cite{PDG2020}, where $n$ is the
number of energy points, $w_i=1/\sigma_i^2$, $x_i$ and $\sigma_i$ are
the measured values and their uncertainties at each energy point,
respectively, and $w=\sum_i w_i$.
 The significance is estimated by comparing the
 likelihoods of the baseline fit and the one defined assuming no
 polarization~\cite{p:value}. The calculation of significance also
 accounts for the systematic uncertainties in the decay parameter,
 mass window, and background. After evaluating the significance
 following each systematic uncertainty, we select the smallest value
 as a conservative estimation. The significance of each energy point is 
 also summarized in table~\ref{summary}.  
Since the significances are all less than $3\sigma$, no evident polarization is observed.
Note that the significance of polarization $P_y$  can not be directly determined by $\Delta\Phi$ and it should be directly determined by $\sin(\Delta\Phi)$  when $P_y \propto \sin(\Delta\Phi)=0$ according to eq.~(\ref{eq:pol}). It although means that $\Delta\Phi$ allows two values either $\Delta\Phi$ or $\Delta\Phi-\pi$, the terms $\sin(\Delta\Phi)$ and $\cos(\Delta\Phi)$ in eq.(\ref{eq:tangles:abcd}) can constraint the unique $\Delta\Phi$ value.
\begin{table}[!htbp]
        %\small
        \caption{The energy spread, $\sigma_{cms}$, the number of observed events, $N_{\rm obs}$, integrated luminosities, $\int {\cal L} dt$,  measured parameters $\eta$, $\Delta\Phi$ and $R$ for each energy point, the weighted average values, and the significance of the $\Delta\Phi$, $S$. For $\sqrt{s}$, the uncertainties are conservative estimates and have the same bias because of the correlation. For $N_{\rm obs}$, the uncertainties are statistical. For measured parameters, the first column uncertainties are statistical and the second column are systematic.}
        \centering
        \scalebox{0.7}{\begin{tabular}{c c c r@{.}l c r c c}
        \hline
        \hline
        $\sqrt{s}$ (MeV) & $\sigma_{cms}$ (MeV) & $\int {\cal L} dt$ ({\rm pb}$^{-1})$	&\multicolumn{2}{c}{$N_{\rm obs}$}	&$\eta$  &\multicolumn{1}{c}{$\Delta\Phi$ (rad)}    &$R$ &$S (\sigma)$\\
        \hline
        $3682.00\pm0.20$ & 1.334  & 404.0&$134$  &$2^{+12.4}_{-11.8}$       &$0.54 \pm 0.17 \pm 0.12$  &$0.38 \pm 0.40 \pm 0.12$   & $0.84 \pm 0.20 \pm 0.14$  &0.9\\
        $3682.75\pm0.20$ & 1.335  & 28.7 &$27$   &$2^{+6.0}_{-5.4}$         &$0.96 \pm 0.13 \pm 0.12$  &$2.35 \pm 1.66 \pm 0.12$   & $0.22 \pm 0.37 \pm 0.34$  &0.5\\
        $3684.22\pm0.20$ & 1.336  & 28.7 &$97$   &$8^{+10.7}_{-10.1}$       &$0.86 \pm 0.15 \pm 0.12$  &$1.19 \pm 0.61 \pm 0.12$   & $0.42 \pm 0.21 \pm 0.20$  &1.8\\
        $3685.26\pm0.20$ & 1.336  & 26.0 &$256$  &$2^{+16.8}_{-16.2}$       &$0.76 \pm 0.10 \pm 0.12$  &$0.16 \pm 0.32 \pm 0.12$   & $0.57 \pm 0.15 \pm 0.16$  &0.3\\
        $3686.50\pm0.20$ & 1.338  & 25.1 &$283$  &$0^{+17.8}_{-16.8}$       &$0.66 \pm 0.12 \pm 0.12$  &$0.02 \pm 0.27 \pm 0.12$   & $0.70 \pm 0.15 \pm 0.15$  &0.0\\
        $3691.36\pm0.20$ & 1.344  & 69.4 &$77$   &$5^{+9.8}_{-8.8}$         &$0.16 \pm 0.23 \pm 0.12$  &$1.29 \pm 0.54 \pm 0.12$   & $1.31 \pm 0.30 \pm 0.16$  &2.8\\
        $3709.76\pm0.20$ & 1.360  & 70.3 &$22$   &$0^{+5.8}_{-4.7}$         &$0.01 \pm 0.40 \pm 0.12$  &$-2.64 \pm 0.60 \pm 0.12$  & $1.55 \pm 0.62 \pm 0.19$  &2.0\\
        AVG.     & $-$ & $-$ &\multicolumn{2}{c}{$-$}             &$0.69 \pm 0.05 \pm 0.05$  &$0.14 \pm 0.16 \pm 0.05$   & $0.72 \pm 0.08 \pm 0.07$  &$-$\\
        \hline
        \hline
        \end{tabular}}
        \label{summary}
\end{table}

\begin{figure}[htbp]
  \centering
  \includegraphics[width=1.0\textwidth]{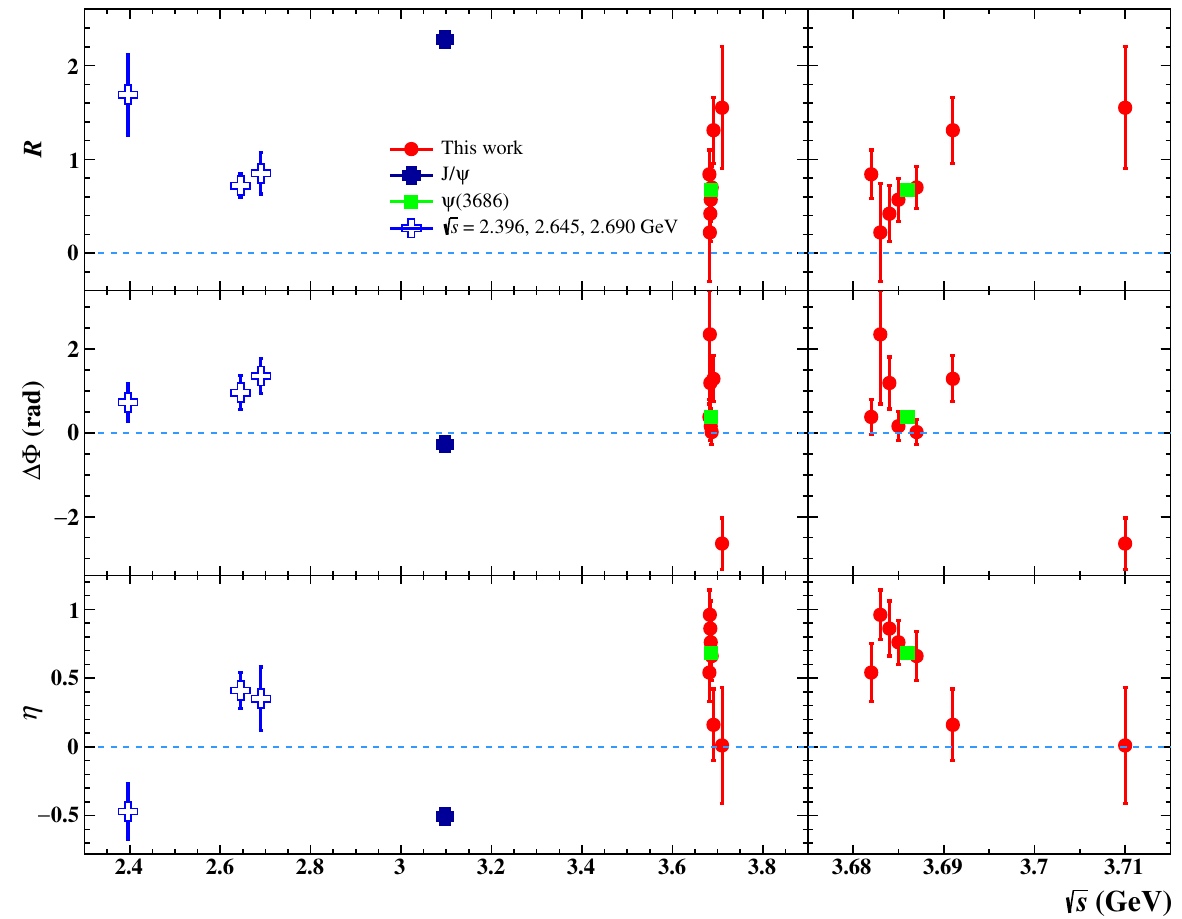}
  \caption{\small The comparisons of $\eta$, $\Delta\Phi$, and $R$ as a
    function of CM energy between this work and the previous
    measurements~\cite{BESIII:2023sgt,BESIII:2023ynq}.}
  \label{dep_ene}
  \end{figure}

\begin{figure}[htbp]
\centering
\includegraphics[width=0.95\textwidth]{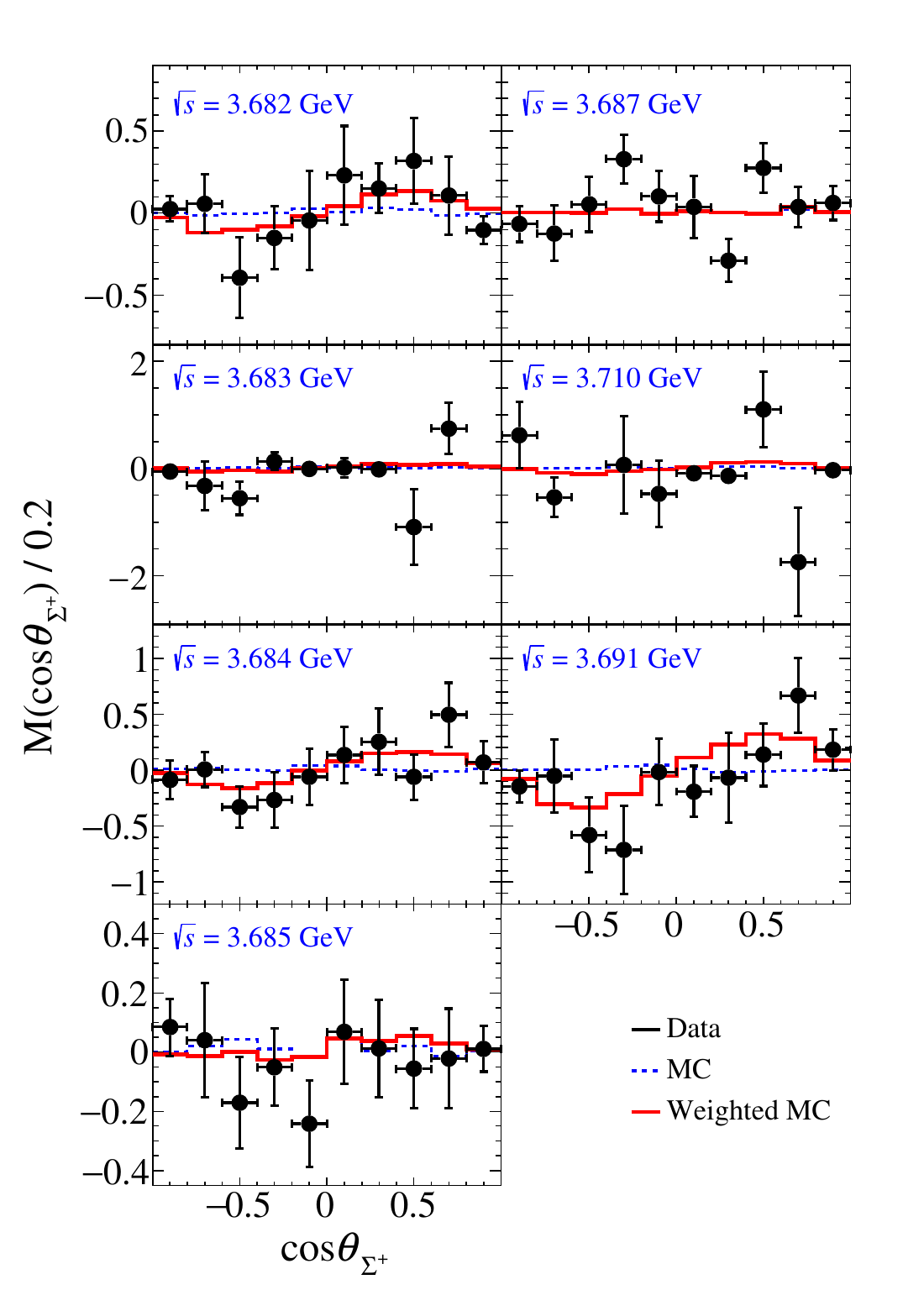}
\caption{\small The moment $M(\cos\theta_{\Sigma^+})$ as a function of
  $\cos\theta_{\Sigma^+}$ for the $\EE\ar\SSB$ reaction at different CM energies. Points with error bars are data, the red solid
  lines are the weighted PHSP MC corrected by the results of global
  fit, and the blue dashed lines represent the distributions without
  polarization from simulated events, evenly distributed in the PHSP.}
\label{cos}
\end{figure}

 The comparison between this work and previous
 measurements~\cite{BESIII:2023sgt,BESIII:2023ynq} is also provided in
 figure~\ref{dep_ene}. Here, $\Delta\Phi$ is less than zero for
 $J/\psi$ decay and $\sqrt{s} = 3.71$ GeV, and greater than zero at
 other energy points, which implies that there may be at least one
 $\Delta\Phi=0$ rad between these energy points. Such an evolution is
 important input for understanding its asymptotic behavior and the
 dynamics of baryons similar to the $\Lambda$
 hyperon~\cite{Mangoni:2021qmd}.
Figure~\ref{cos} shows the distribution of 
the moment given by
\begin{equation}\label{moment}
M(\cos\theta_{\Sigma^+}) = \frac{m}{N}\sum_{i=1}^{N_k}(n^{i}_{1,y} -
n^{i}_{2,y}),
\end{equation}
which is calculated for $m = 10$
intervals in $\cos\theta_{\Sigma^+}$.
 Here, $N$ represents the total
number of events in the data sample.
According to ref.~\cite{BESIII:2022yprl}, $M(\cos\theta_{\Sigma^+})$ is related to the
polarization by
\begin{equation}
M(\cos\theta_{\Sigma^+}) = \frac{\alpha_{\Sigma^+} -
  \alpha_{\bar\Sigma^-}}{2}\frac{1+\eta\cos^2\theta_{\Sigma^+}}{3+\eta}P_y(\theta_{\Sigma^+}).
\end{equation}
Assuming $CP$ conservation, we
have $\alpha_{\Sigma^+} = - \alpha_{\bar{\Sigma}^-}$, and the expected
angular dependence of $M(\cos\theta_{\Sigma^+})$ is
proportional to
%\begin{equation}
$\sqrt{1-\eta^{2}}\alpha_{\Sigma^+}\!\sin\Delta\Phi\cos\theta_{\Sigma^+}\sin\theta_{\Sigma^+}/(3+\eta)$,
which is consistent with the data in figure~\ref{cos}.

\section{Systematic uncertainty}
\noindent
The systematic uncertainties on the measurement of the $\Sigma^+$
hyperon polarization arise due to the $\Sigma^+(\bar{\Sigma}^-)$
reconstruction, the requirements on the $p\pi^0(\bar{p}\pi^0)$ mass
window, the background estimation, the fit method, and the decay
parameters $\alpha_{\Sigma^+/\bar{\Sigma}^-}$.

\subsection{\boldmath $\Sigma^+(\bar{\Sigma}^-)$ reconstruction}
\noindent
The systematic uncertainty due to the $\Sigma^+(\bar{\Sigma}^-)$
reconstruction efficiency, incorporating the efficiencies of tracking,
PID, kinematic fit and requirements of $M_{\pi^0\pi^0}^{\rm Recoil}$
and $\pi^0$ reconstruction, is estimated using the control sample of
$\psi(3686)\ar\SSB$ with the same method as described in
refs.~\cite{BESIII:2021gca,BESIII:2022mfx,BESIII:2024jgy}. The efficiency
difference is evaluated by comparing a full
reconstruction sample that reconstructs both the baryon and anti-baryon
sides, with two partial reconstruction samples that only reconstruct
one side. The MC samples are then corrected with the
efficiency difference, and the fit is repeated. The
differences between the new and nominal values are taken as the
systematic uncertainties.

\subsection{Mass window} 
\noindent
The uncertainty due to the requirements on
the $p\pi^0(\bar{p}\pi^0)$ mass window is estimated with the same
method as introduced in ref.~\cite{BESIII:2020fqg}. The range of the
$p\pi^0(\bar{p}\pi^0)$ mass window is increased or decreased by 1
MeV/$c^2$, and the maximum differences between the new and nominal
values are taken as the systematic uncertainties.

\subsection{Background} 
\noindent
The systematic uncertainty associated with the background is
determined by comparing the results obtained from
the fits with and without considering the sideband background. The
modified likelihood function is expressed as
\begin{equation}
    \mathscr{S} = -\mathrm{ln}\mathscr{L}_{\rm data} + \mathrm{ln}\mathscr{L}_{\rm bg},
\end{equation}
where $\mathscr{L}_{\rm bg}$ is the likelihood function of background events determined in the sideband regions. 

\subsection{Fit method}
\noindent
To validate the reliability of the fit results, an input and output
check is conducted using 500 pseudo-experiments. The helicity
amplitude formula provided in ref.~\cite{Faldt:2017kgy} is
utilized. To ensure that the Gaussian function is fitted
comprehensively, the input values of the polarization parameters are
deliberately chosen to be far away from the fit boundary. In order to
ensure sufficient statistics, 6000 events are generated for each MC
sample. In addition, to avoid the reconstruction effect, the MC
samples used here have not been corrected by the detector. The
differences between the input and output values obtained from the fits
are negligible.

\subsection{Decay parameter}
\noindent
The uncertainties from the decay parameters $\alpha_{\Sigma^+}$ for
$\Sigma^+\to p\pi^0$ and $\alpha_{\bar{\Sigma}^-}$ for
$\bar{\Sigma}^-\to\bar{p}\pi^0$ are estimated by varying the value,
obtained from averaging the results in ref.~\cite{BESIII:2020fqg}, by
$\pm 1\sigma$. The largest differences in the result are taken as the
systematic uncertainties.

\subsection{Total systematic uncertainty}
\noindent
The systematic uncertainties on the polarization measurement are
summarized in table~\ref{uncertainty}. Assuming all sources are
independent, the total systematic uncertainties are determined by adding
these sources in quadrature.

\begin{table}[!htbp]
        %\normalsize
\caption{\small Absolute systematic uncertainties of the measured parameters.}
\centering
\begin{tabular}{l c c}\hline \hline
Source 	& $\eta$   &$\Delta\Phi$ (rad)\\	
\hline
$\Sigma^+ (\bar{\Sigma}^-)$ reconstruction   &0.02     &0.00 \\
Mass window                                  &0.01     &0.02 \\
Sideband background                          &0.11     &0.11 \\
Fit method		                             &0.00     &0.00 \\
Decay parameter                              &0.05     &0.03 \\
\hline
Total                                        &0.12    &0.12 \\
\hline
\hline
\end{tabular}
\label{uncertainty}
\end{table}

\section{Summary}
\noindent
In summary, based on $e^+e^-$ collision data corresponding to an
integrated luminosity of 652.1 pb$^{-1}$ collected with the BESIII
detector at the BEPCII collider, we present an energy-dependent
measurement of transverse polarization, relative phase and the modulus
ratio of the form factors of $\Sigma^+$ hyperon in the $e^+e^- \to
\Sigma^+ \bar{\Sigma}^-$ reaction.  For the first time, the phase of
the $\Sigma^+$ hyperon electromagnetic form factors is explored in a
higher range of four-momentum transfer above
$q^2>13.5~\rm{GeV}^2$. No polarization is evident at each energy point. 
These results are important to
understand the production mechanism of the $\Sigma^+$-$\bar\Sigma^-$
pairs at different energy points.

\newpage
\acknowledgments
\noindent
The BESIII Collaboration thanks the staff of BEPCII and the IHEP computing center for their strong support. This work is supported in part by National Key R\&D Program of China under Contracts Nos. 2020YFA0406400, 2020YFA0406300,  2023YFA1606000; National Natural Science Foundation of China (NSFC) under Contracts Nos. 
12075107, 12247101, 
11635010, 11735014, 11835012, 11935015, 11935016, 11935018, 11961141012, 12025502, 12035009, 12035013, 12061131003, 12192260, 12192261, 12192262, 12192263, 12192264, 12192265, 12221005, 12225509, 12235017; 
the 111 Project under Grant No. B20063; 
the Chinese Academy of Sciences (CAS) Large-Scale Scientific Facility Program; the CAS Center for Excellence in Particle Physics (CCEPP); Joint Large-Scale Scientific Facility Funds of the NSFC and CAS under Contract No. U1832207; CAS Key Research Program of Frontier Sciences under Contracts Nos. QYZDJ-SSW-SLH003, QYZDJ-SSW-SLH040; 100 Talents Program of CAS; The Institute of Nuclear and Particle Physics (INPAC) and Shanghai Key Laboratory for Particle Physics and Cosmology; European Union's Horizon 2020 research and innovation programme under Marie Sklodowska-Curie grant agreement under Contract No. 894790; German Research Foundation DFG under Contracts Nos. 455635585, Collaborative Research Center CRC 1044, FOR5327, GRK 2149; Istituto Nazionale di Fisica Nucleare, Italy; Ministry of Development of Turkey under Contract No. DPT2006K-120470; National Research Foundation of Korea under Contract No. NRF-2022R1A2C1092335; National Science and Technology fund of Mongolia; National Science Research and Innovation Fund (NSRF) via the Program Management Unit for Human Resources \& Institutional Development, Research and Innovation of Thailand under Contract No. B16F640076; Polish National Science Centre under Contract No. 2019/35/O/ST2/02907; The Swedish Research Council; U. S. Department of Energy under Contract No. DE-FG02-05ER41374

\newpage
{\bf \noindent The BESIII collaboration}\\
\\
{\small
M.~Ablikim$^{1}$, M.~N.~Achasov$^{4,c}$, P.~Adlarson$^{75}$, O.~Afedulidis$^{3}$, X.~C.~Ai$^{80}$, R.~Aliberti$^{35}$, A.~Amoroso$^{74A,74C}$, Q.~An$^{71,58,a}$, Y.~Bai$^{57}$, O.~Bakina$^{36}$
% ...
\\
 {\it
$^{1}$ Institute of High Energy Physics, Beijing 100049, People's Republic of China\\
% ...
$^{3}$ Bochum  Ruhr-University, D-44780 Bochum, Germany\\
$^{4}$ Budker Institute of Nuclear Physics SB RAS (BINP), Novosibirsk 630090, Russia\\
% ...
$^{35}$ Johannes Gutenberg University of Mainz, Johann-Joachim-Becher-Weg 45, D-55099 Mainz, Germany\\
$^{36}$ Joint Institute for Nuclear Research, 141980 Dubna, Moscow region, Russia\\
% ...
$^{57}$ Southeast University, Nanjing 211100, People's Republic of China\\
$^{58}$ State Key Laboratory of Particle Detection and Electronics, Beijing 100049, Hefei 230026, People's Republic of China\\
% ...
$^{71}$ University of Science and Technology of China, Hefei 230026, People's Republic of China\\
% ...
$^{74}$ University of Turin and INFN, (A)University of Turin, I-10125, Turin, Italy; (B)University of Eastern Piedmont, I-15121, Alessandria, Italy; (C)INFN, I-10125, Turin, Italy\\
$^{75}$ Uppsala University, Box 516, SE-75120 Uppsala, Sweden\\
% ...
$^{80}$ Zhengzhou University, Zhengzhou 450001, People's Republic of China\\
\\
$^{a}$ Deceased\\
% ...
$^{c}$ Also at the Novosibirsk State University, Novosibirsk, 630090, Russia\\
% ...
}}
%% ends here %%

%\section*{The BESIII collaboration}
%\noindent
%\input{authorlist_2023-12-25}
\end{document}